\documentclass{article}
\pdfoutput=1
\usepackage{arxiv}

\usepackage[utf8]{inputenc} 
\usepackage[T1]{fontenc}    
\usepackage{hyperref}       
\usepackage{url}            
\usepackage{booktabs}       
\usepackage{amsfonts}       
\usepackage{nicefrac}       
\usepackage{microtype}      
\usepackage{lipsum}
\usepackage{framed,multirow}
\usepackage{booktabs}

\usepackage{rotating}
\usepackage{graphicx}
\usepackage{amssymb}
\usepackage{latexsym}

\usepackage{url}
\usepackage{xcolor}

\usepackage{longtable}

\title{A Time-Temperature Dataset for the Strawberry Cold Chain Across Multiple Shipments and Locations}

\author{
 Alla Abdella\thanks{
The authors would like to thank our collaborators for their support of the project including WishFarms for allowing us to conduct the shipping tests and coordinating all the logistics and DeltaTrak for donating the real$-$time loggers used to collect the data in the study.}
\thanks{
This document presents some of the results of the research project funded by the United States Department of Agriculture (USDA) and the Florida Department of Agriculture and Consumer Services (FDACS)[grant number 025792]. Any opinions, findings, and conclusions or recommendations expressed in this material are those of the author(s) and do not necessarily reflect the views of the USDA or FDACS.} \\
  Department of Electrical Engineering\\
 University of South Florida\\
  Tampa, FL 33620 \\
  \texttt{aabdella@usf.edu} \\
   \And
 Jeffrey K. Brecht \\
  Horticultural Sciences Department\\
  University of Florida\\
  Gainesville, FL 32603 \\
  \texttt{jkbrecht@ufl.edu} \\
     \And
 Ismail Uysal \\
  Department of Electrical Engineering\\
 University of South Florida\\
  Tampa, FL 33620 \\
  \texttt{iuysal@usf.edu} \\
}

\begin{document}
\maketitle

\begin{abstract}
This article describes location aware temperature profiles from six strawberry shipments across the continental United States. Three pallets were instrumented in each shipment with three vertically placed loggers to take a longitudinal and latitudinal snapshot of 9 strategically different locations (including the top, middle and bottom layers of the pallets placed in the back, middle and the front of the shipping container) for a combined 54 measurement points across shipments of varying lengths. The sensors were instrumented in the field, right at the point of harvest, recorded temperatures every every 5 to 10 minutes depending on the shipment, and uploaded their data periodically via cellular radios on each device. The data is a result of significant collaboration between stakeholders from farmers to distributors to retailers to academics, which can play an important role for researchers and educators in food engineering, cold-chain, machine learning, and data mining, as well as in other disciplines related to food and transportation.
\end{abstract}


\section{Value of the Data}

\begin{itemize}
\itemsep=0pt
\parsep=0pt
\item With a combined decades worth of research in perishable post-harvest logistics, the authors believe that there is still a lot of unknowns when it comes to what actually happens in the cargo hold throughout the shipment. Another reason why this data is novel and important is that unlike many other studies which involved a singular entity, it is the result of a significant collaboration between all the stakeholders in the cold chain from growers to distributors to retailers to academics.

\item Better understand the temperature profile of a standard cold-chain from harvest to store and how the temperature is distributed inside a perishable produce container during shipments of different durations.

\item We hope that this data will motivate the food transportation research community to delve into developing more sophisticated regression and classification algorithms for univariate and multivariate time series, better clustering methods, learning representations with dimensionality reduction, and a better mathematical and statistical understanding of what is happening in cargo hold.

\item The analysis and processing of temperature time-series data for predictive tasks represent a significant challenge especially when profiles may have variable lengths, high variability, and abnormalities as is common in many cold chain applications. The dataset will enable researchers from a wide array of fields and backgrounds to apply analytical tools such as machine learning and physical models in testing and comparing the performance of their predictive or diagnostic algorithms on the cold-chain.

\item Develop temporal algorithms for solving regression, classification and clustering tasks on time series and deploy advanced data analytics to predict the future behaviour of the data profiles during transportation from harvest to the DC, or to identify if a sensor profile satisfied a specific quality control criteria for a retailer; 

\item Educational purposes which include analysis of univariate or multivariate time series data using statistical methods for regression, deep learning for time series classification, learning representation and location-based prediction; \footnote{Possible sensor and data applications, and several future research trajectories are summarized in \cite{alla2}.}
\end{itemize}

\section{Data Description}
Monitoring and controlling the refrigeration of food during the cold-chain (transportation, storage, and distribution of perishable food items) are critical to reducing the amount of food waste.However, the cost of installation of the monitoring devices such as wireless sensor networks (WSNs) and radio frequency identification (RFID) systems limits monitoring resolution in commercial applications generally to one per container [\cite{JEDERMANN:2009}, \cite{BADIAMELIS:2016}]. Hence, with significant collaboration between the stakeholders from growers to distributors to retailers to academics, six strawberry shipments across the continental United States datasets are shared to help in overcoming this limitation. The datasets now made available were collected aiming at understanding the holistic temperature behavior of the strawberry cold-chain and the development of prediction models to predict the future behavior of the strawberries during transportation from harvest to the DC. Nevertheless, due to the temporal heterogeneity, complexity, similarity, and discrepancy characteristics of the
variables included in these datasets, their use goes beyond this future prediction problem to location-based-prediction, binary control criteria, classification, clustering, etc. \\

\begin{figure}
\centering
  \includegraphics[width=0.9\textwidth]{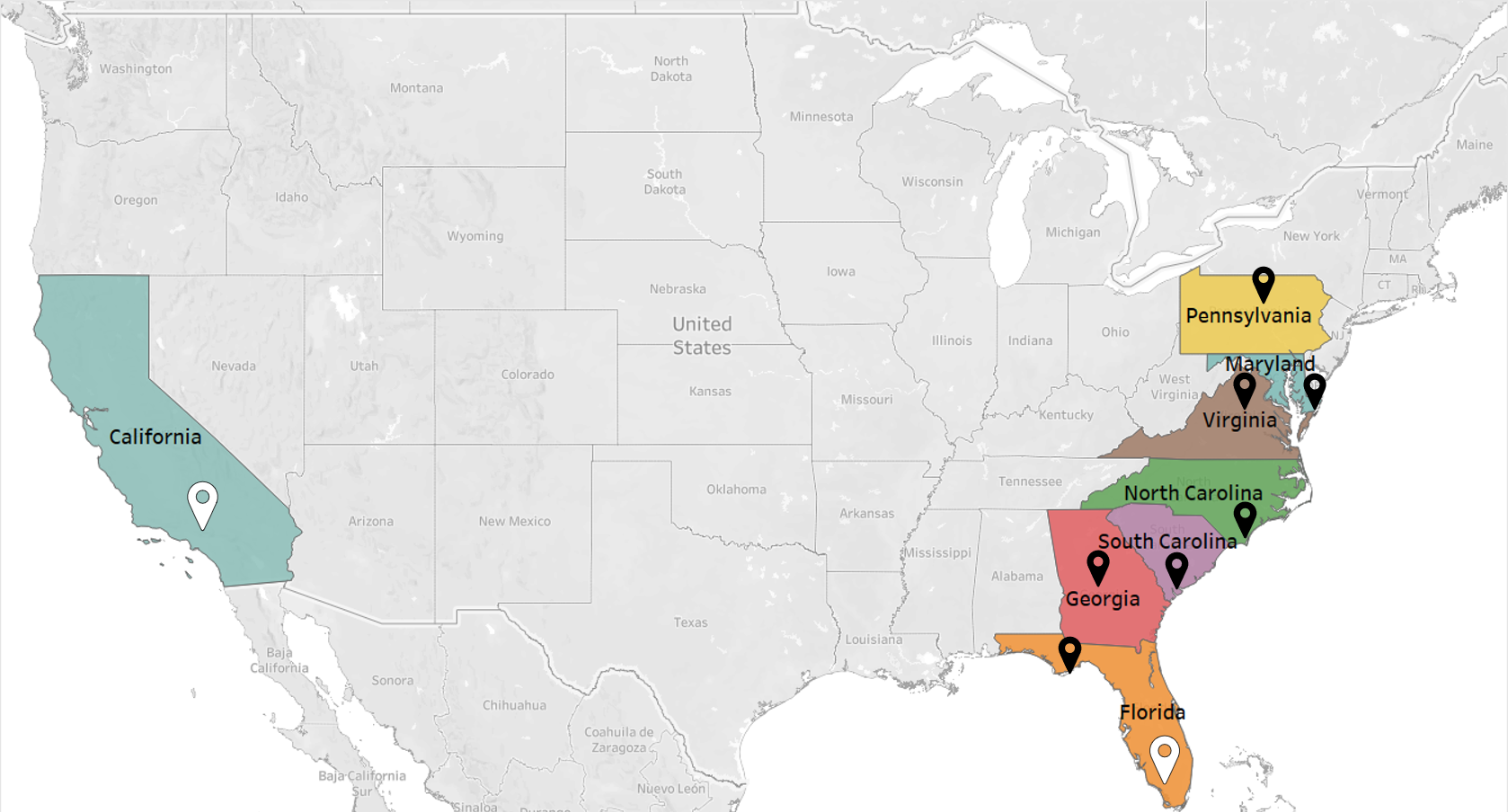}
    \caption{The shipping routes that were monitored in this study:  Two of the shipments originated from Plant City, Florida  with final destinations in Florida and Georgia.  Four shipments originated from Salinas, California with final destinations in Maryland, Pennsylvania, Virginia, South Carolina, North Carolina, Georgia and Texas.}
    \label{FIG:11}
\end{figure}

\section{Specifications} 
\textbf{Subject: }               Agricultural Sciences \\
\textbf{Specific subject area:}   Food engineering and time series\\
\textbf{Type of data:}             Tabular data - CSV files\\
\textbf{Data format}              Mixed (raw and preprocessed) \\

\textbf{How data were acquired:}   Figure \ref{FIG:11} shows a US Map that highlights the shipping routes that were monitored to acquire the data. Two of the shipments originated from Plant City, Florida  with final destinations in Florida and Georgia.  Four shipments originated from Salinas, California with final destinations in Maryland, Pennsylvania, Virginia, South Carolina, North Carolina, Georgia and Texas. \\

\textbf{Instruments:} DeltaTrak’s Reusable Real$-$Time$-$Logger (RTL) Mini devices are used to log both temperature and location data in real time. The RTLs have a wide operational temperature range of $-$30$^{\circ}$C to 95.55$^{\circ}$C with a temperature accuracy of +/$-$1$^{\circ}$C. More information about the hardware used in this study can be found in the appendix. \newline
Data was extracted via the cloud application which can establish secure communications with the GSM loggers. Python \cite{van1995python} was employed to perform subsequent data analysis. \\

\textbf{Parameters for data collection: }             The strawberry cold chain beginning at the field, the strawberries are harvested and placed into plastic clam shells packed into cardboard flats with 8 clamshells (each weighing one pound) per flat, which are subsequently stacked together to build pallets containing between 18$-$20 layers with 6 flats per layer. Once the pallets are built in the field on the back of a flatbed trailer, they are driven to the nearest processing facility to be precooled down to transportation and storage temperatures (0$^{\circ}$C).\\

\textbf{Description of data collection:}            The loggers were instrumented inside the pallets of strawberries right at the point and time of harvest during the pallet buildup stage temperature recording was initiated. A total of three loggers were placed in a single pallet distributed equally along the vertical axis. Specifically one logger was placed closer to the bottom of the pallet (3rd layer from the bottom), another was placed closer to the middle layer of the pallet and a third was placed closer to the top layer of the pallet (3rd layer from the top) between the fruits. A total of three instrumented pallets were sent out with each of the six shipments.  Similar to the placement of loggers within the pallet, the instrumented pallets inside the container were distributed equally along the horizontal axis. Hence, there were 9 loggers in total for each instrumented shipment labeled with respect to the loggers’ location in the pallet and the pallets’ location in the container (front$-$top, middle$-$middle, rear$-$bottom, etc). \newline
The data were transmitted in real time via GSM cellular networks which eliminate the need to collect the loggers at the end of the shipment.\newline

\textbf{Data source location:}   

                         Institution: University of South Florida and University of Florida\newline
                         Country:US\newline
                         States for collected samples/data: 
                         Florida, 
                         Georgia, 
                         Maryland, 
                         Pennsylvania, 
                         Virginia, 
                         South Carolina, 
                         North Carolina, 
                         Texas, 
                         California.
                         
\textbf{Primary data sources:} WishFarms.  \\

\textbf{Data accessibility:}  
                         
                         Direct URL to data:
                       \href{https://data.mendeley.com/datasets/nxttkftnzk/draft?a=7d8b1fed-c1c3-4aa3-8cf3-5b385d221237}{Dataset link }   

\subsection{Related research}

1. \href{https://doi.org/10.1016/j.jfoodeng.2021.110477}{Statistical and temporal analysis of a novel multivariate time series data for food engineering } \cite{alla} ;  

2. \href{https://ieeexplore.ieee.org/abstract/document/9303413}{Sense2Vec: Representation and Visualization of Multivariate Sensory Time Series Data } \cite{alla2}

\section{Variables description}

{

\begin{longtable}{p{22mm} p{10mm} p{68mm}p{40mm} }
\hline
Variable & Type & Description & Source of Data \\
\hline

Front Top  & Numeric &  Sensor placed at the front pallet in the shipping container instrumented closer to the top layer of the pallet (3rd layer from the top) between the fruits. & Shipments 1 through 6 \\

Front Middle  & Numeric & Sensor placed at the front pallet in the shipping container instrumented closer to the middle layer of the pallet between the fruits. & Shipments 1 through 6 \\

Front Bottom  & Numeric & Sensor placed at the front pallet in the shipping container instrumented closer to the bottom of the pallet (3rd layer from the bottom) between the fruits. & Shipments 2 through 6 \\

Middle Top  & Numeric &  Sensor placed at the middle pallet in the shipping container instrumented closer to the top layer of the pallet (3rd layer from the top) between the fruits. & Shipments 1 through 6 \\

Middle Middle  & Numeric & Sensor placed at the middle pallet in the shipping container instrumented closer to the middle layer of the pallet between the fruits & Shipments 1 through 6 \\

Middle Bottom  & Numeric & Sensor placed at the middle pallet in the shipping container instrumented closer to the bottom of the pallet (3rd layer from the bottom) between the fruits. & Shipments 2 through 6 \\

Rear Top  & Numeric &  Sensor placed at the rear pallet in the shipping container instrumented closer to the top layer of the pallet (3rd layer from the top) between the fruits. & Shipments 1 through 6 \\

Rear Middle  & Numeric & Sensor placed at the rear pallet in the shipping container instrumented closer to the middle layer of the pallet between the fruits. & Shipments 1 through 6 \\

Rear Bottom  & Numeric & Sensor placed at the rear pallet in the shipping container instrumented closer to the bottom of the pallet (3rd layer from the bottom) between the fruits. & Shipments 1 through 6 \\
\hline

\end{longtable}
}

\begin{table}
\small
\begin{center}
\scalebox{1}{
\begin{tabular}{llll}
 \multicolumn{4}{c}{} \\
 \hline
Shipments	& 	Approximate Sampling Rate (in Minutes)	&	Number of Samples (per Day)	&	Total Duration (days,hours,minutes)	\\
 \hline

Shipment 1	&	5	&	264	& 6 days, 9 hour and 28 minutes	\\
Shipment 2 &	10	&	127	&	2 days, 1 hour and 9 minutes.	\\
Shipment 3	 &	10	&	132	&	6 days, 9 hours and 25 minutes	\\
Shipment 4	 &	10	&	132	&	5 days, 12 hours and 5 minutes	\\
Shipment 5	& 	10	&	130	&	6 days, 14 hours and 53 minutes	\\
Shipment 6&	10	&	132	&	4 days, 4 hours and 35 minutes	\\

\hline
\end{tabular}}
\caption{Time sampling rate and the duration for all sensor recordings used in the data collection. The calculations include harvest, precooling, and transportation periods. }
\label{tbl5}
\end{center}
\end{table}

\begin{figure*}
\begin{center}

\includegraphics[width=5in]{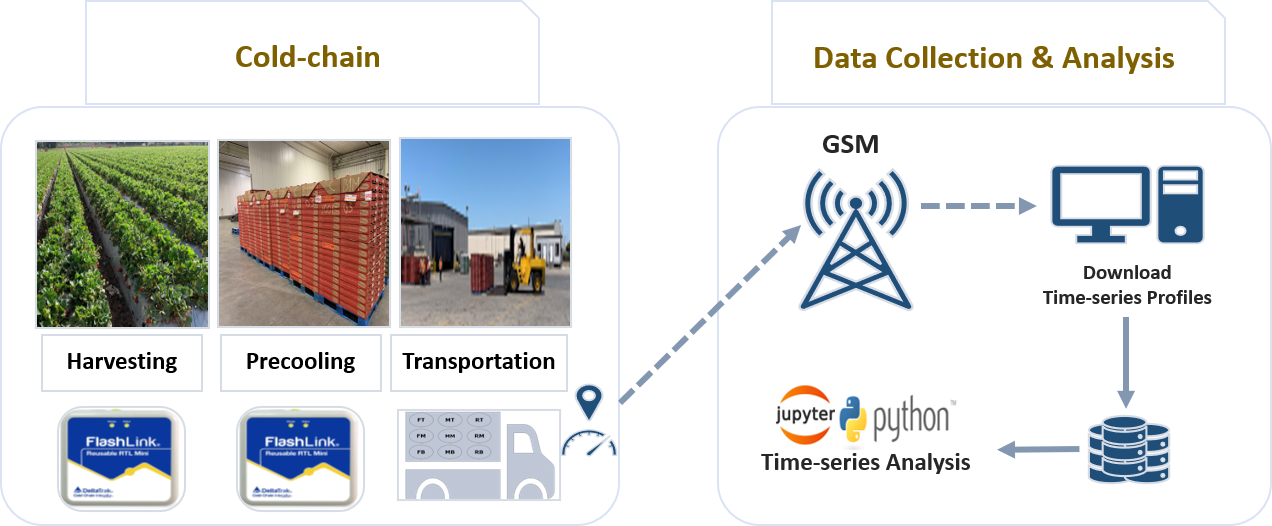}
	\caption{ Cold-chain time-series data collection pipeline. The images show the real environment for data collection in this work.}
   \label{FIG:110}
    
\end{center}
\end{figure*}
\section{Experimental Design, Materials and Methods}

\subsection{Objective}
\begin{figure}
\centering
  \includegraphics[width=.95\textwidth]{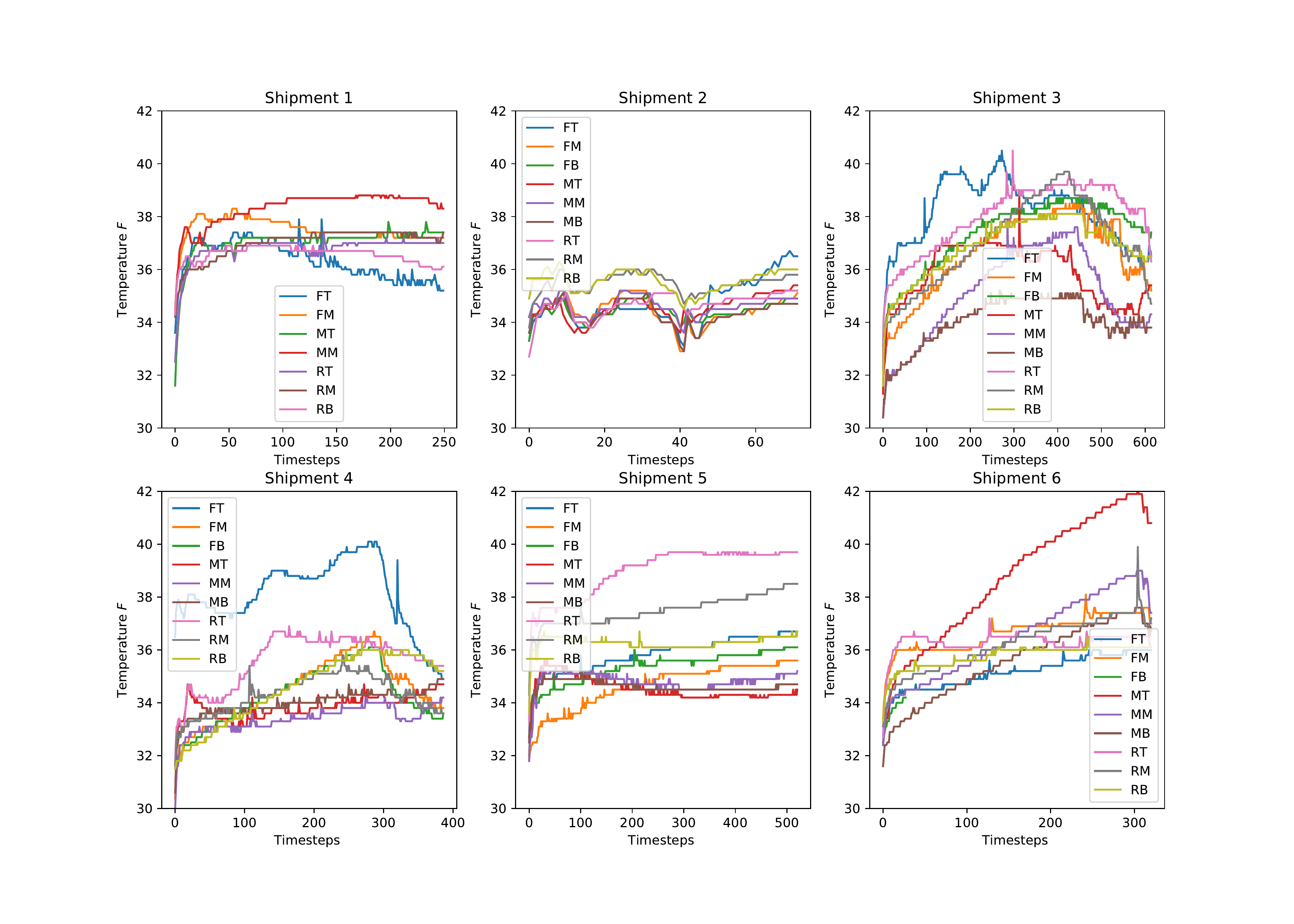}
	\caption{Temperature profiles of multivariate time series data from precooling to the end of transportation.}
	\label{FIG:5001}
\end{figure}

The objective of the data collection process was to obtain the wide range of temperature profiles to which the strawberry shipments around the United States are subjected from the time of the harvest to the arrival at the distribution center and distribution of individual flats to the retail stores as shown by Figure~\ref{FIG:110}. In total there were six shipments as shown by Figure~\ref{FIG:5001}, which cover both short and long$-$distance transportation scenarios (more details are provided in Table 4). Two of the shipments originated from Plant City, Florida with final destinations both in Florida and Georgia.  Four shipments originated from Salinas, California with final destinations in Maryland, Pennsylvania, Virginia, South Carolina, North Carolina, Georgia and Texas. Figure~\ref{FIG:11} highlights the aforementioned states. 

The overall end-to-end data collection stages are highlighted by  Figure~\ref{FIG:110}. The first stage depicts of the beginning of the strawberry cold chain at the field, where they are harvested and placed into clamshells to build pallets. Pallets are then driven to the nearest processing facility to be precooled down to  (0$^{\circ}$C). Finally, the strawberry is transported into different states. The second stage of the pipeline is inspired by a cross-industry standard process for data mining (CRISP-DM)
methodology \cite{Wirth:2000}. There are six phases to consider 1.  Data collection directly from the GSM towers through cloud API. 2. Data description, data quality checking, outliers, and missing values analysis, data exploration, and final preparation; 3. Feature engineering for time series data including scaling, sampling, correlation, and averaging; 4. Machine learning/ Deep learning modeling for time series analysis including representation, forecasting, classification and clustering. 5. Choosing the right evaluation metric based on the problem definition and application (ensures that the model
properly achieves the project objectives); Finally, 6.deploying the learned model into production for practical usage.

\subsection{ Stage 1: Cold Chain}

Traditionally, the strawberry cold chain begins at the field where the strawberries are harvested and placed into plastic clam shells (the packaging the consumers are most familiar with) and packed into cardboard flats with 8 clamshells (each weighing one pound) per flat, which are subsequently stacked together to build pallets containing between 18$-$20 layers with 6 flats per layer. Once the pallets are built in the field on the back of a flatbed trailer, they are driven to the nearest processing facility to be precooled down to transportation and storage temperatures (0$^{\circ}$C). After the precooling is completed, the pallets are either i) shipped directly to the distribution centers (DC) with pending purchase orders from the retailers connected to the DCs or ii) stored at the processing facility if the instantaneous supply from the fields is greater than the accumulated demand from the DCs \cite{Nunes:2014}.

\subsection{ Stage 2: Data Collection}
\begin{figure}[h!]
	\centering
		\includegraphics[width=0.5\textwidth]{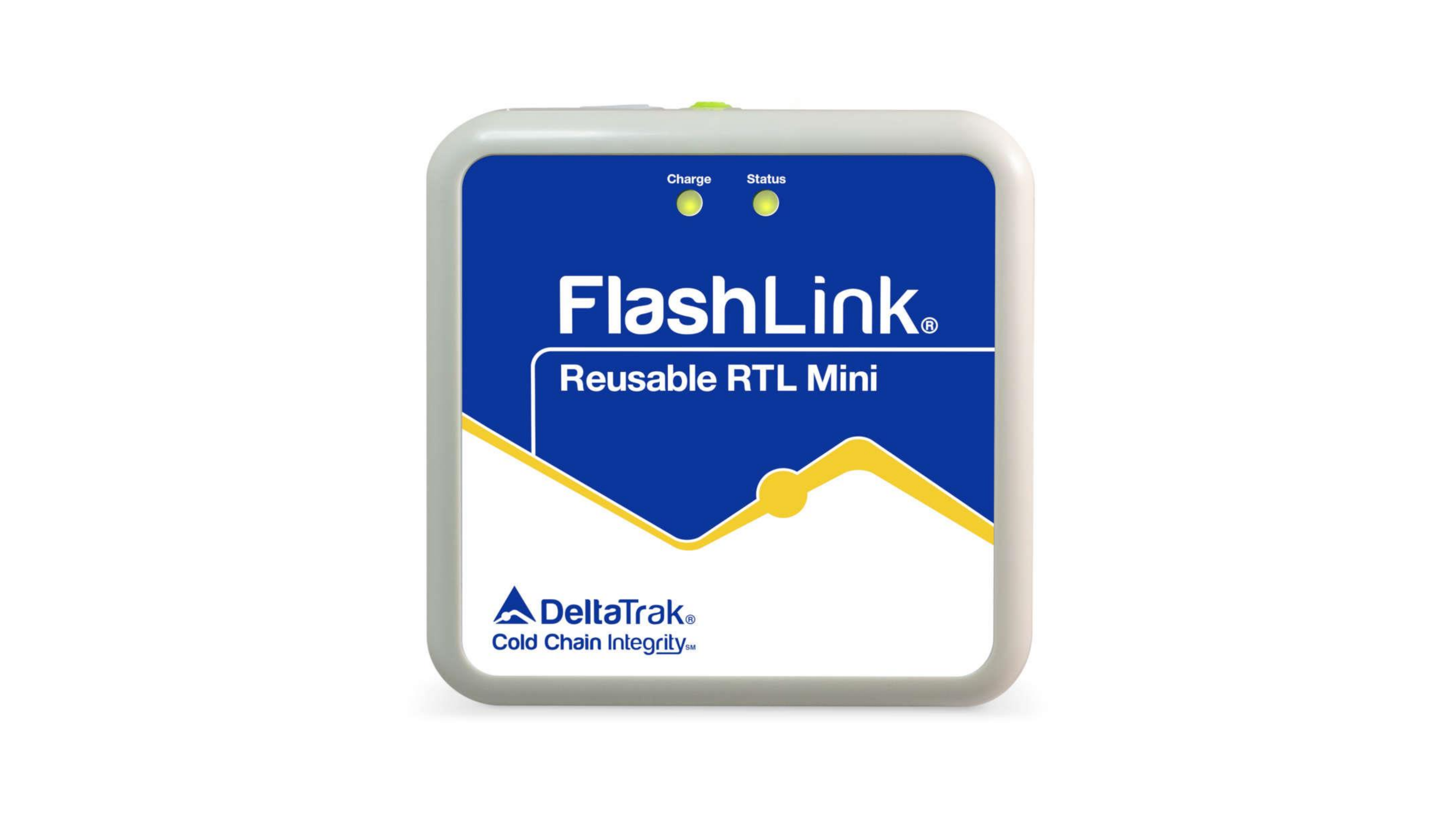}
	\caption{DeltaTrak’s Reusable Real$-$Time$-$Logger (RTL) Mini devices to log temperature data in real time. }
	\label{FIG:10}
\end{figure}

We used DeltaTrak’s Reusable Real$-$Time$-$Logger (RTL) Mini devices as shown in Figure~\ref{FIG:10} to log both temperature and location data in real time.  The data is transmitted in real time via cellular networks which eliminate the need to collect the loggers at the end of the shipment to be able to have access to the recorded data.  The loggers have a wide temperature range of $-$30$^{\circ}$C to 95.55$^{\circ}$C with a temperature accuracy of +/$-$1$^{\circ}$C. The device also eliminates the need for any prior infrastructure setup to be able to automatically collect the data (such as readers for radio frequency identification (RFID) transponders). More information about the hardware used in this study can be found in the appendix. The loggers were instrumented inside the pallets of strawberries right at the point and time of harvest during the pallet buildup stage.  A total of three loggers were placed in a single pallet distributed equally along the vertical axis. A total of three instrumented pallets were sent out with each of the six shipments.  Similar to the placement of loggers within the pallet, the instrumented pallets inside the container were distributed equally along the horizontal axis.  Specifically, one instrumented pallet was placed at the front of the container (i.e., close to the front of the truck), another was placed near the middle of the container and a third was placed at the back of the container (i.e., close to the loading end), each named accordingly (front, middle, rear) in the dataset. Hence, there were 9 loggers in total for each instrumented shipment labeled with respect to the loggers’ location in the pallet and the pallets’ location in the container (front$-$top:FT, front$-$middle:FM, front$-$bottom:FB, ..., rear$-$bottom:RB). Figure~\ref{FIG:5001} displays each sensor profile separately for each of the 6 shipments. Please observe that these figures display real$-$world, noisy and complex multivariate time series as signature representatives of each shipment. Summary statistics for all variables from the shipments datasets are presented in Table 4. These statistics were obtained using the ‘Pandas’ Python package \cite{reback2020pandas}. Python Code used for filtering and analyzing these data will be provided based on request.

\begin{table}
\small
\begin{center}
\scalebox{1}{
\begin{tabular}{ p{2.2cm}p{1.2cm}p{1.4cm}p{0.7cm}p{0.7cm}p{0.7cm}p{0.7cm}p{0.7cm}p{0.7cm}p{0.8cm}p{0.8cm}p{0.8cm}  }
 \multicolumn{9}{c}{} \\
 \hline
Sensor Name	& Ship No &	Timestamps	&	Mean	&	STD	&	Min	&	25\%	&	50\%	&	75\%	&	Max	\\ 

\hline
 & Ship 1	&	250	&	2.40	&	0.39	&	0.89	&	2.11	&	2.39	&	2.72	&	3.28	\\ 
& Ship 2	&	72	&	1.51	&	0.48	&	0.61	&	1.22	&	1.39	&	1.89	&	2.61	\\ 
Front Top & Ship 3	&	615	&	3.43	&	0.67	&	0.61	&	2.78	&	3.61	&	3.89	&	4.72	\\ 
& Ship 4	&	521	&	2.13	&	0.36	&	0.39	&	1.89	&	2.22	&	2.50	&	2.61	\\ 
& Ship 5	&	521	&	2.13	&	0.36	&	0.39	&	1.89	&	2.22	&	2.50	&	2.61	\\ 
&  Ship 6	&	321	&	1.75	&	0.35	&	0.28	&	1.50	&	1.78	&	2.11	&	2.22	\\ 
\midrule
& Ship 1	&	250	&	3.07	&	0.25	&	1.39	&	3.00	&	3.00	&	3.22	&	3.50	\\ 
& Ship 2	&	72	&	1.35	&	0.30	&	0.50	&	1.22	&	1.28	&	1.53	&	1.78	\\ 
Front Middle & Ship 3	&	615	&	2.56	&	0.83	&	-0.39	&	2.00	&	2.78	&	3.28	&	3.72	\\ 
& Ship 4	&	521	&	1.47	&	0.47	&	-0.11	&	1.22	&	1.72	&	1.89	&	2.00	\\ 
& Ship 5	&	521	&	1.47	&	0.47	&	-0.11	&	1.22	&	1.72	&	1.89	&	2.00	\\ 
& Ship 6	&	321	&	2.58	&	0.37	&	0.78	&	2.22	&	2.72	&	2.78	&	3.39	\\ 
\midrule
& Ship 1	&	N/A	&	N/A	&	N/A	&	N/A	&	N/A	&	N/A	&	N/A	&	N/A	\\ 
& Ship 2	&	72	&	1.35	&	0.19	&	0.72	&	1.22	&	1.39	&	1.50	&	1.72	\\ 
Front Bottom &Ship 3	&	615	&	2.99	&	0.70	&	0.39	&	2.61	&	3.22	&	3.50	&	3.72	\\ 
& Ship 4	&	521	&	1.87	&	0.33	&	0.00	&	1.72	&	2.00	&	2.11	&	2.28	\\ 
& Ship 5	&	521	&	1.87	&	0.33	&	0.00	&	1.72	&	2.00	&	2.11	&	2.28	\\ 
& Ship 6	&	28	&	0.93	&	0.26	&	0.28	&	0.72	&	1.00	&	1.11	&	1.22	\\ 
\midrule
& Ship 1	&	250	&	2.79	&	0.37	&	-0.22	&	2.78	&	2.89	&	2.89	&	3.22	\\ 
& Ship 2	&	72	&	1.43	&	0.26	&	0.89	&	1.28	&	1.39	&	1.61	&	1.89	\\ 
Middle Top & Ship 3	&	615	&	2.13	&	0.60	&	-0.39	&	1.61	&	2.39	&	2.72	&	3.89	\\ 
& Ship 4	&	521	&	1.40	&	0.23	&	0.39	&	1.28	&	1.28	&	1.50	&	2.00	\\ 
& Ship 5	&	521	&	1.40	&	0.23	&	0.39	&	1.28	&	1.28	&	1.50	&	2.00	\\ 
& Ship 6	&	321	&	3.74	&	1.31	&	0.22	&	2.72	&	4.00	&	4.89	&	5.61	\\ 
\midrule
& Ship 1	&	250	&	3.51	&	0.36	&	1.28	&	3.39	&	3.72	&	3.72	&	3.78	\\ 
& Ship 2	&	72	&	1.46	&	0.20	&	0.89	&	1.39	&	1.50	&	1.61	&	1.78	\\ 
Middle Middle & Ship 3	&	615	&	1.72	&	0.93	&	-0.89	&	1.11	&	1.72	&	2.61	&	3.28	\\ 
& Ship 4	&	521	&	1.57	&	0.20	&	-0.11	&	1.50	&	1.61	&	1.72	&	1.89	\\ 
& Ship 5	&	521	&	1.57	&	0.20	&	-0.11	&	1.50	&	1.61	&	1.72	&	1.89	\\ 
& Ship 6	&	321	&	2.50	&	0.86	&	0.22	&	1.72	&	2.50	&	3.28	&	3.89	\\ 
\midrule
& Ship 1	&	N/A	&	N/A	&	N/A	&	N/A	&	N/A	&	N/A	&	N/A	&	N/A	\\ 
& Ship 2	&	72	&	1.29	&	0.25	&	0.50	&	1.11	&	1.28	&	1.50	&	1.72	\\ 

Middle Bottom & Ship 3	&	615	&	1.13	&	0.50	&	-0.89	&	0.89	&	1.22	&	1.61	&	1.78	\\ 
& Ship 4	&	521	&	1.45	&	0.12	&	0.28	&	1.39	&	1.39	&	1.50	&	1.78	\\ 
& Ship 5	&	521	&	1.45	&	0.12	&	0.28	&	1.39	&	1.39	&	1.50	&	1.78	\\ 
&Ship 6	&	321	&	1.97	&	0.78	&	-0.22	&	1.39	&	2.11	&	2.72	&	3.11	\\ 
\midrule
& Ship 1	&	250	&	2.65	&	0.30	&	0.28	&	2.61	&	2.78	&	2.78	&	3.00	\\ 
& Ship 2	&	72	&	1.48	&	0.25	&	0.39	&	1.39	&	1.50	&	1.61	&	1.78	\\ 
Rear Top & Ship 3	&	615	&	3.34	&	0.69	&	0.28	&	2.83	&	3.50	&	3.89	&	4.72	\\ 
& Ship 4	&	521	&	3.88	&	0.51	&	0.72	&	3.50	&	4.22	&	4.28	&	4.28	\\ 
& Ship 5	&	521	&	3.88	&	0.51	&	0.72	&	3.50	&	4.22	&	4.28	&	4.28	\\ 
& Ship 6	&	321	&	2.39	&	0.22	&	0.39	&	2.28	&	2.50	&	2.50	&	2.89	\\ 
\midrule
& Ship 1	&	250	&	2.77	&	0.31	&	1.22	&	2.72	&	2.89	&	3.00	&	3.00	\\ 
& Ship 2	&	72	&	1.93	&	0.21	&	1.00	&	1.78	&	2.00	&	2.11	&	2.22	\\ 
Rear Middle & Ship 3	&	615	&	2.79	&	0.91	&	0.11	&	2.11	&	2.78	&	3.61	&	4.28	\\ 
& Ship 4	&	521	&	3.05	&	0.31	&	1.28	&	2.78	&	3.11	&	3.28	&	3.61	\\ 
& Ship 5	&	521	&	3.05	&	0.31	&	1.28	&	2.78	&	3.11	&	3.28	&	3.61	\\ 
& Ship 6	&	321	&	2.34	&	0.54	&	0.61	&	1.89	&	2.50	&	2.72	&	4.39	\\
\midrule
& Ship 1	&	250	&	2.54	&	0.18	&	1.28	&	2.42	&	2.61	&	2.61	&	2.72	\\ 
&Ship 2	&	72	&	1.98	&	0.23	&	1.39	&	1.78	&	2.00	&	2.22	&	2.39	\\ 
Rear Bottom & Ship 3	&	615	&	2.71	&	0.60	&	-0.22	&	2.39	&	2.78	&	3.22	&	3.39	\\ 
&Ship 4	&	521	&	2.37	&	0.13	&	0.89	&	2.28	&	2.39	&	2.50	&	2.61	\\ 
& Ship 5	&	521	&	2.37	&	0.13	&	0.89	&	2.28	&	2.39	&	2.50	&	2.61	\\ 
&Ship 6	&	321	&	2.09	&	0.23	&	0.72	&	1.89	&	2.22	&	2.28	&	2.50	\\ 
\hline
\end{tabular}}
\caption{All sensors across all shipments summary statistics – Integer and numeric variables. }
\label{tbl9}
\end{center}
\end{table}


\begin{table}[h]
\small
\begin{center}
\scalebox{1}{
\begin{tabular}{ p{2.5cm}p{3cm}p{3.5cm}p{3.5cm} }
 \multicolumn{4}{c}{} \\
 \hline
Shipment Name & Variable/Sensor Name	&	Starting Time of Recording	&	Ending Time of Recording	\\
 \hline
	&	Front-Top 	&	3/12/19 12:30 PM	&	4/2/19 12:24 PM	\\
	&	Front-Middle	&	3/12/19 12:28 PM	&	3/22/19 10:19 AM	\\
	&	Front-Bottom	&	NA	&	NA	\\
	&	Middle-Top	&	3/12/19 12:30 PM	&	3/25/19 11:50 PM	\\
\textbf{Shipment 1}	&	Middle-Middle	&	3/12/19 12:26 PM	&	3/22/19 9:27 AM	\\
	&	Middle-Bottom	&	NA	&	NA	\\
	&	Rear-Top	&	3/12/19 12:26 PM	&	4/3/19 8:42 AM	\\
	&	Rear-Middle	&	3/12/19 12:29 PM	&	3/22/19 2:01 AM	\\
	&	Rear-Bottom	&	3/12/19 12:29 PM	&	3/21/19 8:04 AM	\\
\midrule
	&	Front-Top 	&	4/4/19 3:04 AM	&	4/4/19 10:10 PM	\\
	&	Front-Middle	&	4/4/19 3:03 AM	&	4/22/19 7:06 AM	\\
	&	Front-Bottom	&	4/4/19 3:07 AM	&	4/20/19 6:08 PM	\\
	&	Middle-Top	&	4/4/19 3:07 AM	&	4/14/19 3:12 PM	\\
\textbf{Shipment 2}	&	Middle-Middle	&	4/4/19 3:04 AM	&	4/4/19 10:07 PM	\\
	&	Middle-Bottom	&	4/4/19 3:04 AM	&	4/22/19 9:42 AM	\\
	&	Rear-Top	&	4/4/19 3:03 AM	&	4/11/19 7:03 PM	\\
	&	Rear-Middle	&	4/4/19 3:07 AM	&	4/10/19 10:49 AM	\\
	&	Rear-Bottom	&	4/4/19 3:07 AM	&	4/12/19 6:59 AM	\\
\midrule
	&	Front-Top 	&	7/9/19 9:09 AM	&	7/17/19 6:00 PM	\\
	&	Front-Middle	&	7/9/19 8:58 AM	&	7/16/19 3:45 PM	\\
	&	Front-Bottom	&	7/9/19 8:59 AM	&	7/20/19 4:08 PM	\\
	&	Middle-Top	&	7/9/19 8:56 AM	&	7/24/19 6:03 AM	\\
\textbf{Shipment 3}	&	Middle-Middle	&	7/9/19 8:57 AM	&	7/24/19 9:04 AM	\\
	&	Middle-Bottom	&	7/9/19 8:58 AM	&	7/15/19 6:56 PM	\\
	&	Rear-Top	&	7/9/19 8:58 AM	&	7/22/19 3:09 AM	\\
	&	Rear-Middle	&	7/9/19 8:57 AM	&	7/24/19 9:19 AM	\\
	&	Rear-Bottom	&	7/9/19 8:57 AM	&	7/18/19 3:09 AM	\\
\midrule
	&	Front-Top 	&	7/9/19 9:08 AM	&	7/24/19 9:15 AM	\\
	&	Front-Middle	&	7/9/19 8:56 AM	&	7/24/19 8:59 AM	\\
	&	Front-Bottom	&	7/9/19 8:57 AM	&	7/16/19 2:10 PM	\\
	&	Middle-Top	&	7/9/19 9:10 AM	&	7/15/19 3:53 PM	\\
\textbf{Shipment 4}	&	Middle-Middle	&	7/9/19 9:11 AM	&	7/19/19 5:10 AM	\\
	&	Middle-Bottom	&	7/9/19 9:12 AM	&	7/16/19 9:23 AM	\\
	&	Rear-Top	&	7/9/19 9:12 AM	&	7/24/19 9:19 AM	\\
	&	Rear-Middle	&	7/9/19 9:13 AM	&	7/19/19 5:00 AM	\\
	&	Rear-Bottom	&	7/9/19 9:13 AM	&	7/16/19 2:34 PM	\\
\midrule
	&	Front-Top 	&	7/10/19 10:24 AM	&	7/17/19 5:19 PM	\\
	&	Front-Middle	&	7/10/19 10:24 AM	&	7/24/19 3:10 PM	\\
	&	Front-Bottom	&	7/10/19 10:24 AM	&	7/20/19 1:47 PM	\\
	&	Middle-Top	&	7/10/19 10:19 AM	&	7/18/19 11:01 AM	\\
\textbf{Shipment 5}	&	Middle-Middle	&	7/10/19 10:32 AM	&	7/19/19 7:07 AM	\\
	&	Middle-Bottom	&	7/10/19 10:41 AM	&	7/18/19 11:35 AM	\\
	&	Rear-Top	&	7/10/19 10:22 AM	&	7/18/19 4:31 PM	\\
	&	Rear-Middle	&	7/10/19 10:33 AM	&	7/18/19 5:21 PM	\\
	&	Rear-Bottom	&	7/10/19 10:22 AM	&	7/24/19 2:06 PM	\\
\midrule
	&	Front-Top 	&	7/10/19 10:22 AM	&	7/26/19 5:03 AM	\\
	&	Front-Middle	&	7/10/19 10:18 AM	&	7/25/19 10:19 AM	\\
	&	Front-Bottom	&	7/10/19 10:39 AM	&	7/10/19 8:32 PM	\\
	&	Middle-Top	&	7/10/19 10:25 AM	&	7/18/19 9:51 AM	\\
\textbf{Shipment 6}	&	Middle-Middle	&	7/10/19 10:25 AM	&	7/15/19 6:29 AM	\\
	&	Middle-Bottom	&	7/10/19 10:25 AM	&	7/16/19 11:06 AM	\\
	&	Rear-Top	&	7/10/19 10:26 AM	&	7/17/19 7:58 PM	\\
	&	Rear-Middle	&	7/10/19 10:27 AM	&	7/17/19 5:55 PM	\\
	&	Rear-Bottom	&	7/10/19 10:27 AM	&	7/23/19 4:43 AM	\\

\hline
\end{tabular}}
\caption{Time intervals for the temperature measurements for all sensor recordings used in the data collection.  }
\label{tbl14}
\end{center}
\end{table}

\clearpage

\appendix
\section{Supporting information}
Supplementary data associated with this article can be found in the online version at \href{https://data.mendeley.com/datasets/nxttkftnzk/draft?a=7d8b1fed-c1c3-4aa3-8cf3-5b385d221237}{Supplemental files}  \newline  \\

\appendix
\section{Real Time Logger (RTL) Mini Devices}
Please visit the following website for more information about the sensors used in collecting the data: 
\url{https://www.deltatrak.com/reusable-real-time-loggers}


\bibliographystyle{unsrt}  
\bibliography{manuscript}  


\end{document}